\documentclass[aip,rsi,reprint,superscriptaddress,floatfix]{revtex4}

\usepackage{amsmath,latexsym,amssymb,wasysym}
\usepackage{float}
\usepackage[dvipdfm]{graphicx}
\usepackage{epsfig}
\usepackage[usenames]{color}

\newcommand{\um}{\ensuremath{\mu \mathrm{m}}}

\begin{document}

\title{Development of a confocal rheometer for soft and biological materials}
\author{S.~K.~Dutta}
\author{A.~Mbi}
\author{Richard~C.~Arevalo}
\author{Daniel~L.~Blair}
\affiliation{Department of Physics, Georgetown University, Washington,
DC 20057, USA}

\begin{abstract}
We discuss the design and operation of a confocal rheometer, formed by
integrating an Anton Paar MCR301 stress-controlled rheometer with a
Leica SP5 laser scanning confocal microscope.  Combining two
commercial instruments results in a system which is straightforward to
assemble that preserves the performance of each component with
virtually no impact on the precision of either device.  The instruments are
configured so that the microscope can acquire time-resolved,
three-dimensional volumes of a sample whose bulk viscoelastic
properties are being measured simultaneously.  We describe several aspects of the design and, to demonstrate the system's capabilities, present the results of a few common measurements in the study of soft materials.

\end{abstract}


\maketitle

\section{Introduction}

Complex fluids exhibit unique mechanical properties that are
determined by the physicochemical details of their constituent
components. One common example is a colloidal dispersion, where nano-
to micrometer sized particles are suspended within a fluid -- {\em
  e.g.} milk, paint and blood are all colloidal dispersions. The {\em
  structuring} of the fluid has two predominant effects: an
enhancement of the fluid viscosity and, when the concentration of the
dispersed material is high enough, the appearance of a frequency
dependent elastic modulus. These characteristics are determined using
rheology, the measurement science of quantifying the response of
fluid based materials to an applied stress or strain.

Rheology techniques are classified into two main types known as bulk
and micro, each having specific benefits and limitations. Bulk
measurements often require relatively large sample volumes and,
depending on the measuring system, can suffer from a somewhat limited
range due to the inherent inertia of the tooling. The main advantage
of bulk rheology is that the highly nonlinear mechanical behavior of
soft materials can be directly determined. Passive microrheology
utilizes the energy spectrum of thermal fluctuations and provides very
localized structural information over a tremendous dynamical range
that is limited only by the acquisition rate of the
measurement.\cite{Dasgupta:2002uw} The primary drawbacks of
microrheology are understanding the implications of incorporating
tracer particles and the limitations set by the magnitude of the
thermal fluctuations; effectively, microrheology is limited to nearly
homogeneous systems with very small elastic moduli. Overall, rheology
is extremely powerful as a characterization tool for a broad class of
biologically derived or chemically synthesized materials. However, in
many instances where bulk- and micro- rheology are applied,
information about the role of structure, either inherent or influenced
by boundary conditions, are essentially unknowable. Limited access to
structure results in a great deal of uncertainty about the microscopic
origins of the mechanical response.

The first instruments specifically developed for optically quantifying
the structural response of complex fluids to an externally applied
shear stress were based on X-ray and neutron scattering.
\cite{Panine:2003bq, Sasa:2010hb,Porcar:2011gc} Scattering methods are
particularly powerful for investigating average structural changes,
such as conformational changes in protein
networks\cite{Weigandt:2009je} and the bulk phase behavior of
worm-like micelles.\cite{Liberatore:2006dq} If the material is
inherently disordered, which is the case for most soft and biological
materials, scattering can only provide spatially averaged information
that generally precludes details about localized structural
rearrangements driven by thermal excitations and external stresses.  A
natural extension of the scattering approach is to directly measure
the real space structural response of complex fluids under shear
through the use of optical microscopy. This need has driven the
development of new optical-rheology platforms with ever increasing
sophistication and versatility.\cite{Bender:1995dr, Cohen:2004hp, Besseling09a, Basu11a, Paredes11a, Cheng11a, Boitte:2013dn} Access to time-resolved,
three-dimensional information is crucial for an accurate
quantification of the microscopic structure that ultimately determines
material properties; connecting macroscopic observables, such as shear
and bulk moduli, to the relevant physical interactions and structure
is a cornerstone of modern materials science.

The instrumentation challenge remains clear and open: provide a
measurement device that combines high resolution, high magnification,
real space, time resolved spatial information in three dimensions that
is coupled with simultaneous high resolution mechanical deformations
that can be easily reproduced by research groups. Precision
measurements of the structure and mechanical properties of soft
materials requires sub-micrometer spatial resolution and
nanonewton-meter torque resolution. Luckily, devices that
independently attain these levels of precision are commercially
available.

The technology of fast laser scanning and spinning disk confocal
microscopy techniques (LSCM or SDCM) has matured dramatically over the
past twenty years. Innovations made to confocal microscopy, through
commercial and academic partnerships, are providing unprecedented
gains in imaging resolution at ever increasing acquisition speeds and
at steadily decreasing costs.\cite{Hell:2003if} Confocal microscopy
has emerged as a powerful tool in soft materials
physics,\cite{Lu:2008fo} as it provides three-dimensional
reconstructions of structures at sub-micrometer resolution. The
principles of confocal microscopy are straightforward; by
discriminating out of focus light, sharp two-dimensional images are
``stacked'' in the third dimension, providing time resolved volumetric
data. These image volumes can then be rapidly analyzed at the sub-voxel level using advanced processing techniques.\cite{Crocker96a, MPT}
There are many commercial implementations from all of the major
microscopy companies, and a number of component built systems that are
generally based on spinning disk platforms.

Concomitant to the advances of confocal microscopy, stress-controlled
rheometer technology has also advanced, providing new standards for
sensitivity and stability. Moreover, stress-controlled rheometers are
ideally suited as development platforms due to the combination of low
friction, feedback-controlled, inductive motors that provide precise
torque and position encoders that measure displacements. In nearly
all implementations, the the motor/encoder systems are integrated into
the {\em upper} tool. This compact design is in contrast to strain-controlled platforms where one tool rotates to provide displacement,
while the other tool responds to the stress that propagates through
the sample. The single tool configuration provides the flexibility to
incorporate versatile modifications of the static portion of the lower
tooling.

What follows below is a detailed description of what is needed to
reproduce our confocal rheometer system. A key component of our plan
was to produce a functional device within a very short period of time
with a limited amount of machine work. To attain this does require the
use of a particular rheometer (Anton Paar MCR series) and therefore
emulating this system with other devices may prove difficult.  We do
feel that this system can be reproduced using devices from other
manufacturers if the specific guidelines we provide are
transferred. We will discuss a number of design criteria that guided
our development and a series of data that will help motivate the
intended demand.  We also quantify a few unavoidable limitations of
these systems and confirm some of the capabilities through recent
publications.\cite{Schmoller10a, Arevalo11a, Holland12a}

\section{Design Principles}

The plates of shearing devices should remain parallel at a fixed gap
throughout their entire range of motion, whether that corresponds to a
full rotation for a rheometer or a maximum displacement for a linear
shear stage.  The limitations of these devices are set by the
tolerances attainable through computer numerical control (CNC)
machining methods; in most instances, CNC machining can attain $2\ \um$ precision.  Most modern rheometers are produced within these
specifications and therefore can reproducibly attain gaps to within
$20\ \um$ while remaining functional and allowing easy interchange
between tools.  Tool runout and parallelity are dramatically
compromised if the relative orientation of each tool is not maintained
for all applications, leading to a reduction in the reliability of
reported rheological data.  Therefore, when developing a new rheological system, either with a top-down design using existing technology or a bottom-up design by assembling custom components, each new
implementation must at least match these specifications.

We have chosen a top-down approach for constructing a confocal
rheometer.  The primary considerations for this decision were time and
functionality.  Having a working device within a year, from conception
to implementation, was highly desirable and attainable.  By using
devices with guaranteed factory specifications, we could forgo years of
engineering and benchmarking.  Furthermore, commercial instruments
potentially provide a more user friendly system, making training,
operation, and collaboration simpler.

\begin{figure}
  \includegraphics[width=3.0in]{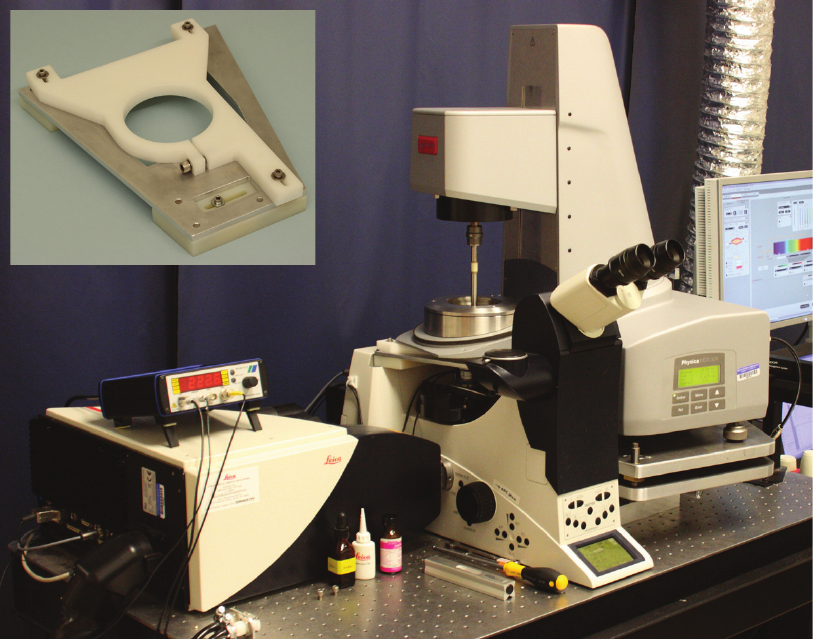}
  \caption{\label{FPhoto} Photograph of the confocal rheometer.  The
    base plate of an Anton Paar MCR301 rheometer is replaced with a
    metal cup.  Optical access for a Leica SP5 confocal microscope is
    provided by a glass coverslip mounted in the cup, which serves as
    the rheometer bottom plate.  The field of view of the microscope
    can be changed by moving the rheometer on a manual three-axis
    translation stage.  The inset shows the device that clamps the cup to the microscope stand to reduce vibrations.}
\end{figure}

Our system, shown in Fig.\ \ref{FPhoto}, consists of an MCR301 stress-controlled rheometer from Anton Paar GmbH and an SP5 LSCM from Leica
Microsystems; it was designed in collaboration with both companies.
Optical access to the sample for the microscope is provided from below
by a glass coverslip which also serves as the bottom plate for the
rheometer.  This coverslip is rigidly mounted to the rheometer via a
metal cup; as a result, the two devices maintain autonomous
functionality even when joined together.

The rheometer modifications made for this application do not dictate a
choice of microscope manufacturer or confocal head style.  However, if
an attempt is made to duplicate this design using another rheometer
manufacturer, care must be taken to match the mounting and tool
specifications.  Maintaining gap tolerance and tool runout are
critical features for simultaneous rheology and visualization.  We
feel that if a microscope stage or any other independent platform is
substituted for the bottom plate of the rheometer, the task of
maintaining a gap on the order of $50\ \um$ across a tool diameter of
$25\ \mathrm{mm}$ becomes dramatically more difficult. In our design,
we are able to machine fixed components at tolerances that are within
the manufacturers specifications, providing us reliable gaps of $h \ge
20 \mu$m.

\section{System Components}

\subsection{Rheometer}

The rheometer for the system was modified at the factory by Anton Paar
to be compatible with our design.  This primarily involved the relocation
of the front control panel and removal of the lower front section of
the rheometer; this provides unimpeded access below the lower tool
platform.  The result is a horizontal platform that accepts all
standard and custom manufactured bottom plate accessories.  These
modifications, and the large distance from the tool rotational axis to
the front of the rheometer body, provide sufficient clearance for the
microscope when the two instruments are mounted next to each other.

\begin{figure}
  \includegraphics[width=3.0in]{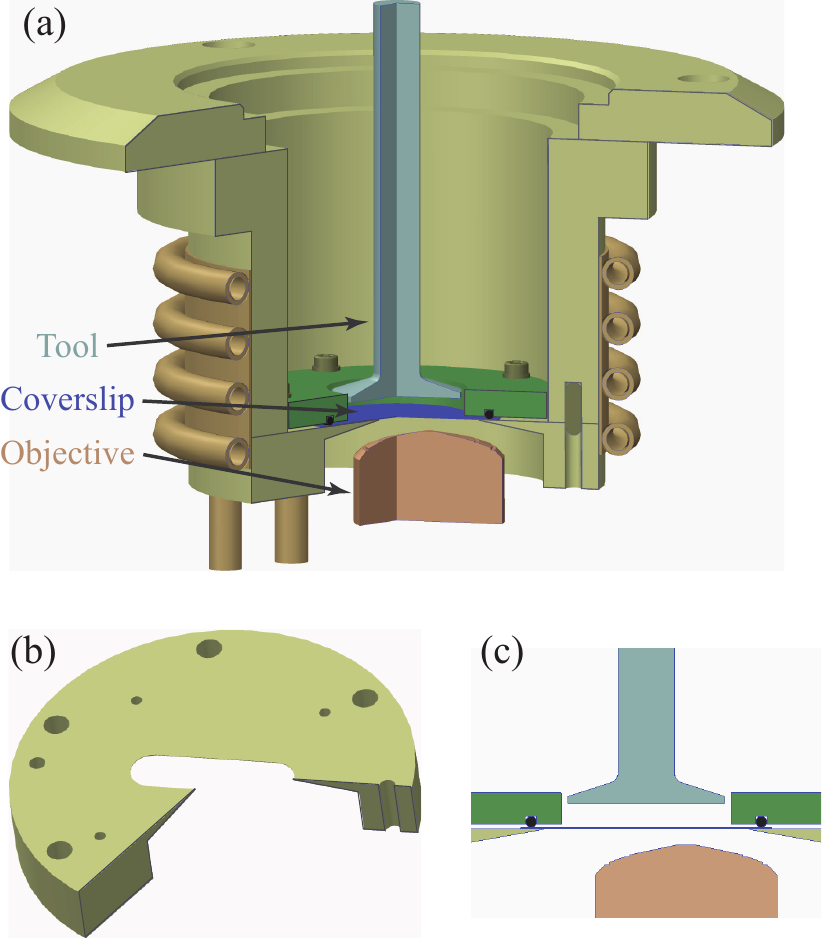}
  \caption{\label{FDrawings} Cut-away drawings of the (a) confocal
    rheometer assembly (including the microscope objective, metal cup
    that mounts to the rheometer, glass coverslip, and rheometer tool)
    and (b) the baseplate at the bottom of the cup.  (c) Magnified
    cross section of the sample region.}
\end{figure}

A custom stainless steel cup bolts to the rheometer platform and
positions the bottom plate coverslip so that the platform does not
interfere with the microscope; see Fig.\ \ref{FDrawings}(a).  The cup
consists of a factory-supplied mounting flange, a cylindrical section
that bolts to the flange, and an interchangeable baseplate that bolts
to the cylindrical section.  The baseplate, an example of which is
drawn in Fig.\ \ref{FDrawings}(b), registers the coverslip and has a
slotted opening to provide optical access for the microscope.  The
underside of the baseplate is machined to a sharp edge around the
perimeter of the slot; this allows a high magnification immersion
objective to reach the coverslip, thus preserving its full working
distance.  A flat acrylic ring, fitted with a rubber O-ring, clamps
the circular coverslip (40 mm in diameter) in place.  A
cross-sectional view of the sample region is shown in
Fig.\ \ref{FDrawings}(c).

We use standard measuring system tools from Anton Paar.  They must, however,
be 150 mm in length in order to reach the coverslip.  The cup assembly
can accommodate tools, in either plate or cone geometries, with
diameters of up to 25 mm.

The cup can be covered to limit evaporation from aqueous samples and
is water tight so that biological samples can be immersed in media.  A
bath circulator flows heated or chilled fluid through copper coils
that wrap around the cup to regulate the temperature of the entire cup
and sample.

\subsection{Microscope}
\label{SScope}

The Leica confocal uses a standard DMI6000B inverted microscope with
the upper illumination arm removed.  The vertical ($z$) position of the objective
is controlled by a piezo-based focusing attachment from piezosystem jena.  It
is important to take into account the further geometrical constraints
that the body of the piezo puts on the design of the rheometer
cup. Alternatively, motor control of the nosepiece will suffice, but
will lead to slower acquisition times for confocal stacks.

Images are acquired with a raster point scanner, which can be operated
in an 8 kHz resonant mode.  The acquisition speed can be effectively
doubled with a bi-directional $x$ scan; in this case, a typical image
stack with a resolution of $256 \times 256 \times 100$ voxels can be
acquired in roughly 2 s.  In addition, a series of three-dimensional stacks can be
imaged in a bi-directional $z$ mode, where the order of $z$ slices is
reversed on alternate stacks, preventing the piezo from resetting
abruptly.

\subsection{Translation Stage and Instrument Coupling}

To adjust the position of the rheometer relative to the microscope
body, and thus the imaging field of view, we designed a mounting stage
that provides leveling and translation capabilities.  The stage is
comprised of two aluminum plates separated by three fine-threaded,
ball-bearing-tipped screws for leveling the rheometer.  These plates
are held together with three springs.  The springs are removable and
provide access to the bottom plate which bolts directly to two
orthogonal Edmund Optics 38-180 translation stages.  The translation
stages are then mounted to a plate that bolts to the breadboard of the
air table (Technical Manufacturing Corporation, model 63-543) on which
the entire confocal rheometer sits.  In order to place the rheometer's
tool over the objective, the stage must be rotated about it's vertical
center-line axis at an angle of $12.5^\circ$ relative to the the
microscope -- {\em N.B.}\ this is the case for Leica 6000 series
inverted microscopes, however each different manufacturer should be
tested for orientation requirements.

Through the use of a vibrationally isolated table, mechanical noise
from external sources is diminished for the microscope and rheometer
separately.  However any residual relative motion between the
instruments can degrade the imaging quality.  The biggest source of
this motion comes from the active components of the rheometer itself,
with dominant frequencies near 100 Hz.  To minimize relative motion, a
``soft-coupling'' clamp, shown in the inset of Fig.\ \ref{FPhoto}, connects the microscope base and the rheometer
cup.  This clamp is adjustable in the
horizontal plane, so the rheometer can be positioned as needed, and is
attached to the microscope stand through the stage mounting positions.
While the vibrational noise with the clamp (with an
amplitude under 100 nm, as directly measured from rapidly acquired microscope images) is still larger than when using a
traditional microscope stage, the current conditions do not hinder the
sort of measurements described in the next section.

\begin{figure}
  \includegraphics[width=3.0in]{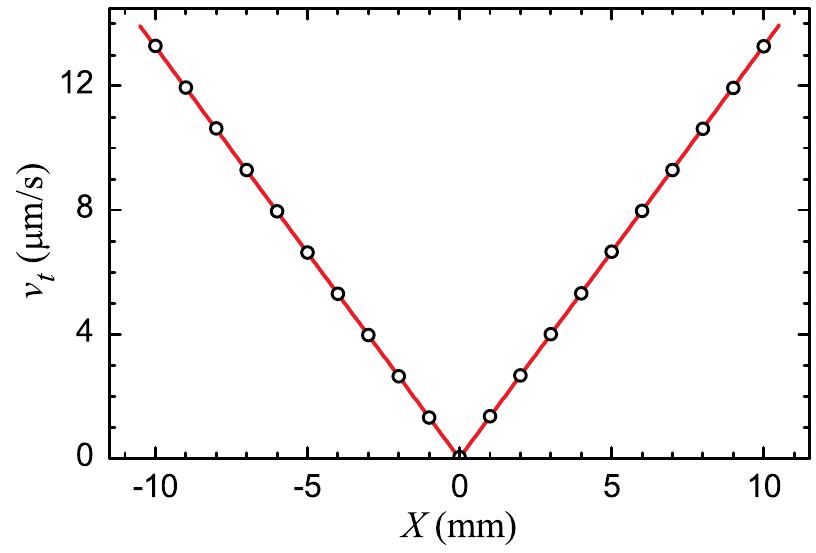}
  \caption{\label{FRCalibrate} Radial position calibration.  The
    measured linear shear velocity $v_t$ (symbols) of the rheometer tool
    displays the expected linear dependence (solid line) on the
    position $X$ of the rheometer stage.}
\end{figure}

Placing the rheometer on its translation stage only roughly constrains
its position, which simplifies assembly of the instrument.  To
accurately determine the radial position of the microscope objective
with respect to the central axis of the rheometer tool, we rely on
imaging the surface of the tool directly.  For instance, this position can be
determined by measuring the local linear velocity $v_t$ of the tool due
to a rotation of known angular velocity.  We verified this approach
with the measurement shown in Fig.\ \ref{FRCalibrate}.  Images of the
tool were acquired in reflectance while the rheometer was set to a
steady shear rate of 0.330 1/s with a gap of $50\ \um$.  We
measured $v_t$ at several locations across the face of the tool as the
stage was moved in one dimension.  As this path was chosen to run
through the rheometer axis, $v_t$ should have a linear dependence on the
stage position $X$.  The fit shown with a solid line in Fig.\ \ref{FRCalibrate} yields a shear
rate of 0.331 1/s, where the discrepancy from the nominal value
reflects the size of the mismatch in the spatial and temporal
calibrations of the two instruments.

\section{Experimental Results}

To demonstrate the capabilities of the instrument as well as provide
some context for the discussion of other design issues, we next
present a few representative measurements.  Throughout the section,
the local velocity, vorticity, and gradient axes at the imaging location will be referred to as $x$, $y$, and $x$, respectively.

\subsection{Oscillatory Measurements}

Many biological polymer networks have complex rheological properties
that play important roles in structural integrity and cell motility.
Furthermore, individual fiber bundles can often be imaged, opening up
the possibility of linking bulk behavior and various geometrical
properties of these sparse disordered networks.  We now show a simple
example of how the viscoelasticity of a gel, as quantified by the dynamic
shear moduli, is reflected in its structure.

\begin{figure}
  \includegraphics[width=3.0in]{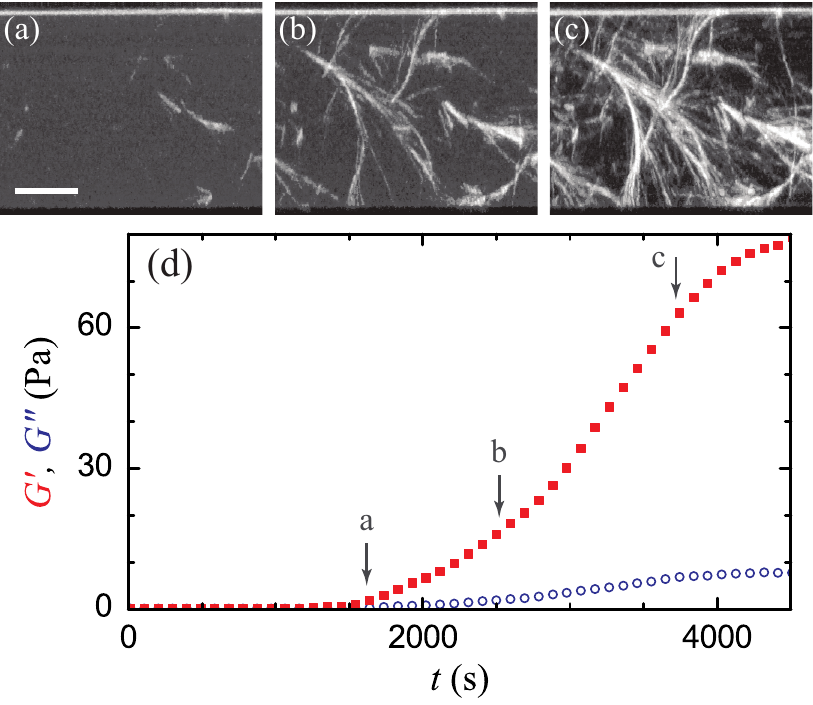}
  \caption{\label{FCollagen} Polymerization of collagen.  (a)-(c) The images show snapshots in the $xz$-plane of a sample during polymerization; the scale bar indicates $30\ \um$.  (d) The storage $G^\prime$ (squares) and loss $G^{\prime\prime}$ (circles) moduli both plateau as the gel is formed.  The arrows indicate the times at which the images were acquired.}
\end{figure}

Figure \ref{FCollagen} shows data taken during the polymerization of a fluorescently-labeled collagen network (1 mg/mL concentration, 0.12 ionic strength).  Once the polymerization was
initiated and the sample was loaded, the rheometer gap was set to
$90\ \um$ with a 25 mm parallel plate tool.  The storage $G^\prime$ and loss $G^{\prime\prime}$ moduli were continuously monitored using
a 0.5\% strain amplitude, 1 Hz oscillation.  The results, shown in Fig.\ \ref{FCollagen}(d), suggest that the polymerization took roughly 4000 s.

At the same time, three-dimensional image stacks were periodically acquired at a radial position of 8 mm.  Figure \ref{FCollagen}(a)-\ref{FCollagen}(c) show images in the $xz$-plane taken at the times indicated by the arrows in the rheology plot.  Each panel is the maximum projection along the $y$-axis of a section of the stack $9\ \um$ thick and shows the full extent of the sample in $z$, from coverslip to tool.  From these images, where brighter colors indicate a stronger fluorescence signal, the formation of fibers can be followed.  It is interesting to note that the background noise signal diminishes as more fluorescently-labeled collagen is incorporated into the network.

For this measurement, the polymerization was carried out on a clean
glass coverslip.  However, if adhesion is a concern for larger
strains, we have found it possible to chemically treat the coverslips
to make them hydrophilic (or hydrophobic, when needed)
without affecting the imaging quality in any appreciable way.

\subsection{Steady Flow Measurements}

Many colloidal systems and other structured fluids display interesting
behavior under a continuous shear.  This may include ordering at the level
of individual constituent particles or the formation of large-scale
shear bands.  The confocal rheometer can provide some insight into how
these structural properties affect bulk rheology under a wide range of
shears.

\begin{figure}
  \includegraphics[width=3.0in]{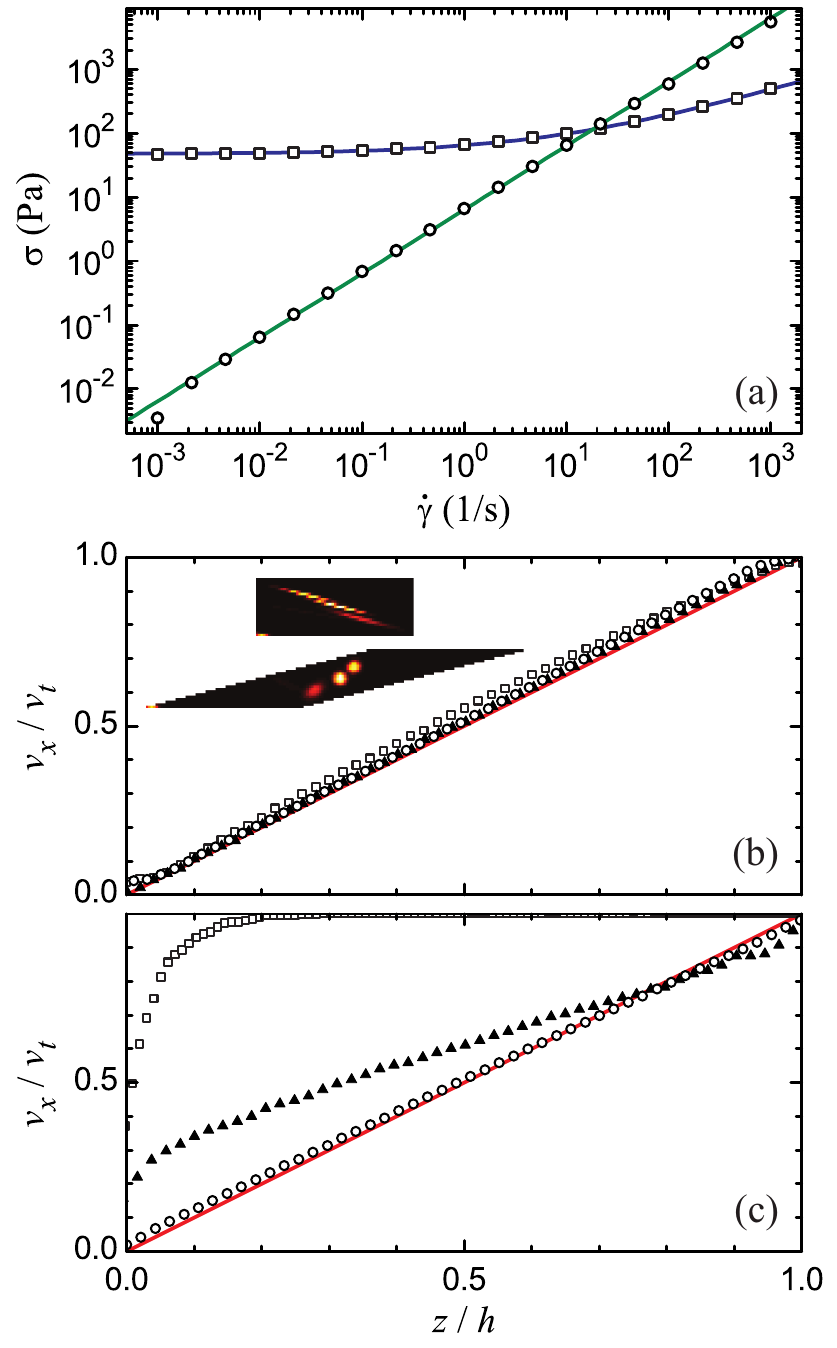}
  \caption{\label{FFlow} Flow properties of two fluids.  (a) Rheological flow curves for honey (circles) and a compressed emulsion
    (squares) show Newtonian and Herschel-Bulkley behavior, respectively.  The $z$-dependent flow profiles for (b) honey and (c) the emulsion
    have a very different dependence on the shear rate; here, the
    average shear velocity $v_x$ (normalized by the local tool speed
    $v_t$) is plotted for local shear rates of $10^{-3}$ (squares), 1
    (triangles), and $10^3$ (circles) $1 / \mathrm{s}$.  The images
    inset in (b) show an example of the analysis used to extract $v_x$
    for the fastest rate.}
\end{figure}

A comparison of the flow behavior for honey, a nearly Newtonian fluid,
and an oil-in-water emulsion (compressed to a volume fraction of
0.60), which has a yield stress and shear thins, is shown in
Fig.\ \ref{FFlow}.  A clean coverslip worked well for the honey, but this surface lead to complete boundary slip for the emulsion drops.  We found that a robust solution was to lithographically define a square grid of posts on a coverslip using SU-8, a negative photoresist that adheres well to glass.  Image degradation can be avoided by matching the index of refraction of the posts to that of the sample.  As there was also considerable slip on the bare metal rheometer tool, we attached a similarly roughened coverslip to its surface.

The difference in the bulk behavior of the materials is clearly
seen in the flow curves of Fig.\ \ref{FFlow}(a).  The shear stress
$\sigma$ is nearly linear in the strain rate $\dot{\gamma}$ for the
honey (circles), corresponding to a constant viscosity of 6.4 Pa s.
On the other hand, the emulsion (squares) closely follows a
Herschel-Bulkley form, as shown by the solid fit line, with $\sigma =
47 + 17 \dot{\gamma}^{0.47}$.

A simple way to characterize the spatial properties of a steady state
flow is with the average velocity $v_x$ in the shear direction as a
function of the position $z$ above the coverslip.  Such flow profiles are shown in Fig.\ \ref{FFlow}(b)
and (c) for the two fluids for three different shear rates.  The velocities are normalized by the local tool speed $v_t$; the position is normalized by the rheometer gap $h$, which was set to $100\ \um$.  The flow was measured by following 1 \um\ fluorescent tracer
beads mixed in with the samples.

As seen in Fig.\ \ref{FFlow}, the honey undergoes a roughly affine deformation (indicated by the solid line) for all $\dot{\gamma}$, as expected in
a flow where the shear stress is independent of $z$.  The emulsion
flows in a similar fashion for the highest shear rate, but displays
strong shear localization near the coverslip at the lowest shear rate.

The advanced engineering of both the confocal and rheometer are needed
to acquire data over the wide dynamic range shown in the figure.  For
instance, the torque needed by the rheometer to produce the shears in
Fig.\ \ref{FFlow}(a) for the honey varies by six orders of magnitude.

In terms of imaging, for the slow rate, the samples move slowly enough
for full three-dimensional stacks to be acquired over time.  With these
in hand, traditional particle finding and tracking algorithms can be
used to extract three-dimensional velocity fields from the individual tracers.\cite{Crocker96a, MPT}  At higher shear, where it is no
longer possible to acquire full stacks, two-dimensional images can be rapidly
acquired at fixed $z$.  From these images, particle image velocimetry
techniques yield values of $v_x$ and $v_y$ averaged over the
$xy$-plane.

At still higher shear, even a single image is distorted by the raster
scanning process, as the fluid moves a significant amount in the time
it takes to acquire one line.  In this case, a single velocity
component of a flow, assumed to be uniform in the direction of
scanning, can be inferred from the relative displacement
between pairs of image lines needed to recover circular particles.
This type of analysis was used for the highest shear rates for both fluids; an example
of the image processing is shown in the inset to Fig.\ \ref{FFlow}(b).
The raw image is at the top, while the one below it shows the recovered image due
to an average flow of $6700\ \um / \mathrm{s}$.  The frequency of the resonant
scanner limits the velocity to which this technique can be applied.

When acquiring a three-dimensional stack with a confocal microscope, as needed for a
flow profile measurement, it can be critical to know the exact value
of $z$ at which each slice is taken.  As there are several changes in the index of refraction along the imaging path, knowledge of the objective
position is insufficient.  In particular, the index mismatch between
the immersion fluid (with index $n_i$) and the sample (with $n_s$)
introduces a number of effects, including reduced signal intensity, a
degradation in resolution, and a shift in the position of the focal plane.\cite{Carlsson91a}

\begin{figure}
  \includegraphics[width=3.0in]{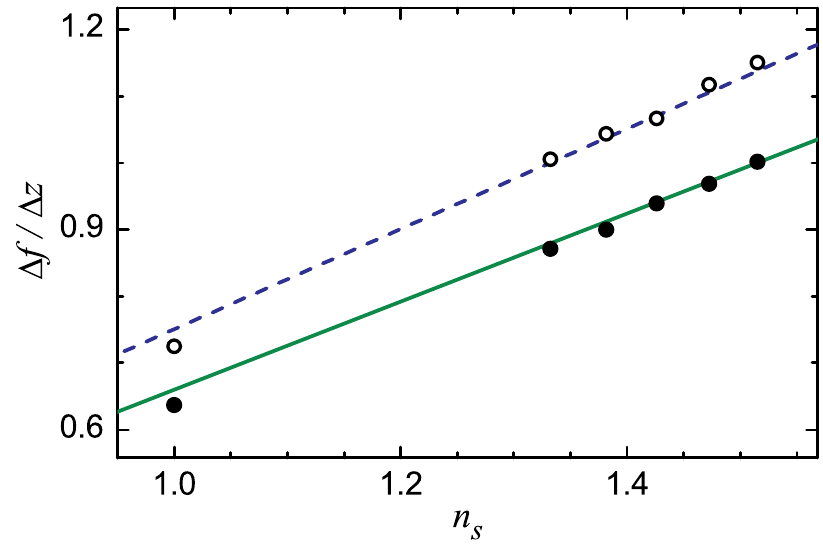}
  \caption{\label{FFocalShift} Focal shift measurements.  The ratio $\Delta f / \Delta z$ between changes in the position of the focal plane and the objective increases roughly linearly with the index of refraction $n_s$ of the sample being imaged.  Measurements are shown for two objectives which use oil (solid circles) and water (open circles) as the immersion fluid.}
\end{figure}

We measured the focal shift for two objectives and a selection of samples that filled the rheometer's gap, as shown in Fig.\ \ref{FFocalShift}.  For each sample, a linear dependence was found between the rheometer's gap and the position of the objective needed to bring the surface of the tool into focus. The focal shift can be quantified by the slope $\Delta f / \Delta z$ of this relationship, where a change in the objective height $\Delta z$ leads to the focal plane moving by $\Delta f$.  As expected, the values are nearly reproduced by $n_s / n_i$ for two $63\times$ objectives using oil (solid circle, solid line) and water (open, dashed) immersion fluid.  This correction was applied to the flow profiles in Fig.\ \ref{FFlow}; in fact, the water-immersion objective had to be moved 89 and $95\ \um$ for the honey and emulsion, respectively, to cover the actual $100\ \um$ gap.  While the focal shift can be measured with far less instrumentation, the confocal rheometer is well suited to the task, particularly given the ease of setting a variable gap, and can provide a value for any specific combination of objective and sample.

Performing the flow measurements presented a few other challenges,
particularly for the emulsion.  A consequence of providing optical
access to the sample is that the coverslip that serves as the bottom
rheometer plate can deflect over the area where it is unsupported by
the metal baseplate.  This can occur in two ways.  For one, loading a
stiff sample can result in a significant force on the glass
originating from the tool.  Additionally, for objectives that require
an immersion fluid, the resultant coupling can lead to a deflection of
the coverslip when the objective moves in the $z$-direction.  This
problem is exacerbated by the presence of the rheometer tool, which
imposes a fixed boundary plane.  If the coverslip moves for either
reason, it imposes a stress on the sample which can result, for
example, in the rearrangement of emulsion droplets.

There are several ways to mitigate these issues.  To minimize
deflection, a rheometer cup baseplate with a single small hole [rather
  than one with a wide slot, as show in Fig.\ \ref{FDrawings}(b)] can
be used to reduce the unsupported coverslip area.  Using water or a
low viscosity oil as the immersion fluid greatly reduces the objective
coupling.  Ill effects of the coupling can be further reduced with the
bi-directional $z$ scanning described in Sec.\ \ref{SScope}, which is
essential for imaging colloidal systems.

The severity of the deflection depends strongly on the properties of
the sample, so it is important to monitor its effects.  In particular,
we have found that the normal force sensor of the rheometer is
sensitive to very small deflections of the coverslip, including those
caused by the objective during sequential three-dimensional stack acquisition.  When
loading a stiff sample, the rheometer control software can lower the
tool at a slow enough rate to allow the sample to fill the gap
uniformly.  In addition, for a structured fluid that displays yielding
behavior, a slow rotation or oscillation of the tool can improve this
relaxation.  In practice, we modify the loading profile to minimize
the coverslip deflection, as measured by the normal force sensor and
direct imaging with the microscope.

\begin{figure}
  \includegraphics[width=3.0in]{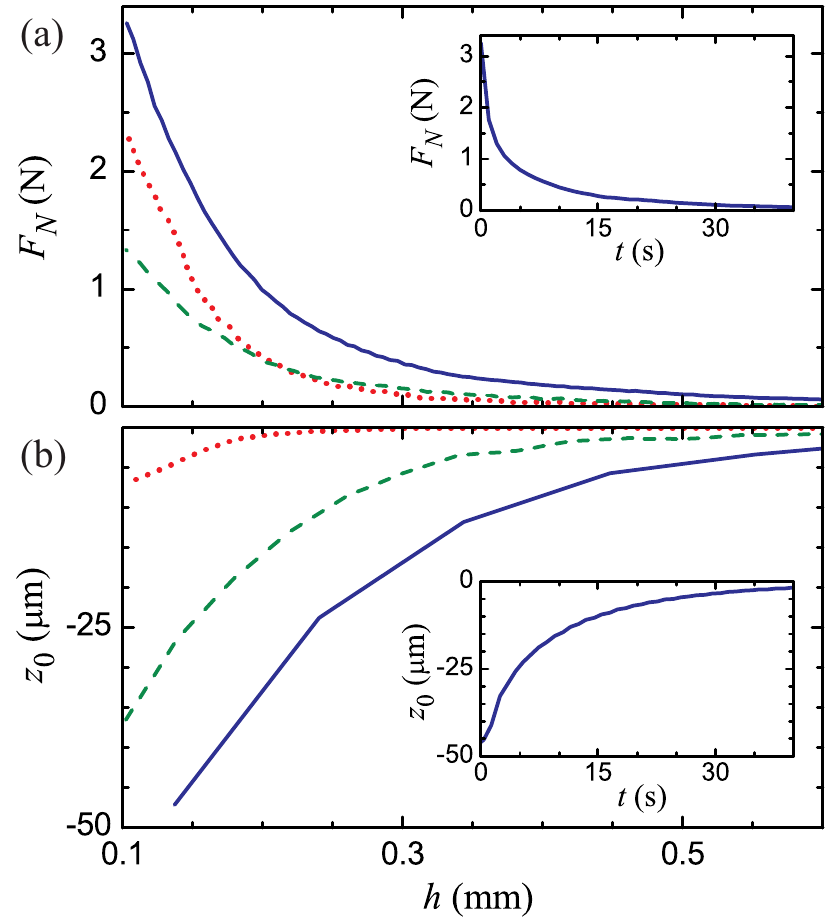}
  \caption{\label{FLoading} Sample loading.  The (a) normal force and (b) coverslip deflection were monitored while a compressed emulsion was loaded under the different conditions described in the text.  In the main plots, the gap $h$ decreases with time to a final value of 100\ \um.  The insets show the relaxation of the coverslip when the tool was rotated after loading, as a function of time $t$.}
\end{figure}

An example of this monitoring is shown in Fig.\ \ref{FLoading}, where the normal force $F_N$ and coverslip location $z_0$ (where $z_0 = 0$ before the sample is loaded) were measured during the loading of a compressed emulsion to a gap $h$ of $100\ \um$.  Both of these values were quite high when the tool was lowered at a rate of $50\ \um/\mathrm{s}$ (solid lines).  As shown in the insets, the coverslip relaxes quickly when the tool is rotated at a shear rate of 5 1/s after the gap has been set to its final value.  The maximum normal force is smaller when loading at $5\ \um/\mathrm{s}$ with a 1 1/s rotation (dashed lines).  Alternatively, the deflection can be minimized by using a baseplate with a single hole (dotted lines).  This baseplate also limits deflection during measurements after the sample has been loaded.  Thus the baseplate and loading protocol can be chosen to meet the needs of each experiment.

\section{Summary}

We have described a confocal rheometer comprised of two commercial
instruments.  While the Anton Paar rheometer did have to be modified
to provide adequate access for the Leica microscope, taking advantage
of the significant engineering of the instruments simplified assembly
of the combined system and went a long way to ensuring successful
operation.  In addition to providing standard bulk viscoelastic
measurement capability, the rheometer has many advantages over a
standard shear cell, including a normal force sensor that is
particularly useful during loading, easy measurement profile
definition, a wide range of applied torque, and a gap that can be
precisely controlled over the course of an experiment.

We feel that due to the large number of groups that have incorporated
microscopy and rheology as equipment for their research, that this
system can be implemented with a modest amount of additional machine work
and engineering. We hope that our straightforward design principles
will be easily transferable.

\section{Acknowledgments}

We are indebted to L.~Der for his insight on a variety of design
issues and machining expertise.  This work was funded by the generous
support of Georgetown University and the National Science Foundation
Grant \# DMR-0847490.


\begin{thebibliography}{21}
\bibitem[{MPT()}]{MPT}
\bibinfo{note}{Matlab versions of the tracking code are available at
  physics.georgetown.edu/matlab}.
\end{thebibliography}
\end{document}